\def\rfr#1{eq. (\ref{#1})}

\def\bar{\begin{eqnarray}}
\def\ear{\end{eqnarray}}
\def\bb{\bibitem}
\def\eqi{\begin{equation}}
\def\eqf{\end{equation}}
\def\eqia{\begin{eqnarray}}
\def\eqfa{\end{eqnarray}}
\def\rp#1#2{{#1\over#2}}

\def\lb#1{\label{#1}}

\def\olt{\dot\omega_{\rm GM}}
\def\oge{\dot\omega_{\rm 1PN}}
\def\ogge{\dot\omega_{\rm 2PN}}
\def\oms{\dot\omega_{\rm meas}}

\def\psr{PSR J0737-3039A/B\ }

\def\oc2{$\mathcal{O}(c^{-2})$}

\documentclass{elsart5p}
\usepackage{ifpdf}
\usepackage{graphicx,natbib,amssymb,lineno}
\ifpdf
\usepackage[%
  pdftitle={Instructions for use of the document class
    elsart},%
  pdfauthor={Simon Pepping},%
  pdfsubject={The preprint document class elsart},%
  pdfkeywords={instructions for use, elsart, document class},%
  pdfstartview=FitH,%
  bookmarks=true,%
  bookmarksopen=true,%
  breaklinks=true,%
  colorlinks=true,%
  linkcolor=blue,anchorcolor=blue,%
  citecolor=blue,filecolor=blue,%
  menucolor=blue,pagecolor=blue,%
  urlcolor=blue]{hyperref}
\else
\usepackage[%
  breaklinks=true,%
  colorlinks=true,%
  linkcolor=blue,anchorcolor=blue,%
  citecolor=blue,filecolor=blue,%
  menucolor=blue,pagecolor=blue,%
  urlcolor=blue]{hyperref}
\fi

\makeatletter
\def\elsartstyle{%
    \def\normalsize{\@setfontsize\normalsize\@xiipt{14.5}}
    \def\small{\@setfontsize\small\@xipt{13.6}}
    \let\footnotesize=\small
    \def\large{\@setfontsize\large\@xivpt{18}}
    \def\Large{\@setfontsize\Large\@xviipt{22}}
    \skip\@mpfootins = 18\p@ \@plus 2\p@
    \normalsize
}
\@ifundefined{square}{}{}
\makeatother

\journal{New Astronomy}
\begin{document}

\begin{frontmatter}
\title{Prospects for measuring the moment of inertia of pulsar J0737-3039A}

\author{Lorenzo Iorio\corauthref{boh}}
\address{Istituto Nazionale di Fisica Nucleare (INFN), Sezione di Pisa. Address for correspondence: Viale Unit$\grave{a}$ di Italia 68, 70125\\Bari (BA), Italy}

\corauth[boh]{Corresponding author}
\ead{lorenzo.iorio@libero.it}
%\ead[url]{authors.elsevier.com/locate/latex}

\begin{abstract}
Here we consider the possibility$-$envisaged by many authors as feasible in the near future$-$of measuring at $10\%$ or better the moment of inertia $I$ of
the pulsar J0737-3039A  via the gravitomagnetic spin-orbit periastron precession (analogous to the Lense-Thirring pericentre precession in the case of a test-particle orbiting a central spinning mass).
Such a gravitomagnetic effect is expected to be of the order of $10^{-4}$ deg yr$^{-1}$ and the present-day precision in measuring the periastron precession of J0737-3039A via pulsar timing  is $6.8\times 10^{-4}$ deg yr$^{-1}$. However the  systematic uncertainty in the much larger first-order post-Newtonian (1PN) gravitoelectric  precession (analogous to the Einstein Mercury's perihelion precession in the weak-field and slow-motion approximation), which should be subtracted from the measured one in order to pick up the gravitomagnetic rate, is of primary importance. Indeed, by determining the sum of the masses by means of the third Kepler law,
 such a bias amounts to $0.03165$ deg yr$^{-1}$, according to the current level of accuracy in knowing the parameters of the  J0737-3039 system. The major sources of uncertainty  are  the Keplerian projected semimajor axis $x_{\rm B}$ of the component B and the post-Keplerian parameter $s$, identified with $\sin i$; their knowledge should be improved by three orders of magnitude at least; the bias due to the  Keplerian projected semimajor axis $x_{\rm A}$ of the component A amounts to $\approx 10\%$ today. The present-day level of accuracy in the eccentricity $e$ would affect the investigated measurement at a percent level, while the impact of the orbital period $P_{\rm b}$ is completely negligible. If, instead, the sum of the masses is measured by means of the post-Keplerian parameters $r$ and $s$, it turns out that $r$ should be measured five orders of magnitude better than now: according to the present-day level of accuracy, the total uncertainty in the 1PN periastron rate is, in this case,  $2.11819$ deg yr$^{-1}$. In conclusion, the prospect of measuring the
moment of inertia of PSR J0737-3039A at 10$\%$ accuracy or better seems
unlikely given the limitations to the precision with which the
system's basic binary and post-Keplerian parameters can be measured via
radio timing.

 %In conclusion, the
%prospects for obtaining the required precision are bad.

 %Our analysis fully confirms the statements made in the past by other authors concerning the necessity of having three parameters measured at high %accuracy.
  \end{abstract}

\end{frontmatter}

\section{Introduction}
 The  measurement of the moment of inertia $I$ of a neutron star at a $10\%$ level of accuracy or better would be of crucial importance for effectively constraining the Equation-Of-State (EOS) describing  matter inside neutron stars \citep{Mor04,Bej05,Lat05,Lav07}.

 After the discovery of the double pulsar \psr system \citep{Bur03}, whose relevant orbital parameters  are listed in Table \ref{tavola}, it was often argued that such a measurement for the A pulsar via the post-Newtonian gravitomagnetic spin-orbit periastron precession \citep{Bar75a,Dam88,Wex95} would be possible after  some years of accurate and continuous timing.
%
% %
%
% %
%
%
\citet{Lyn04} write: ``Deviations
from the value predicted by general relativity may be caused by contributions from spin-orbit
coupling \citep{Bar75b}, which is about an order of magnitude larger than for PSR B1913+16. This
potentially will allow us to measure the moment of inertia of a neutron star for the first time \citep{Dam88, Wex95}.''

According to \citet{Lat05}, ``measurement of the spin-orbit perihelion advance seems possible.''

In \citep{Kra06} we find:  ``A future determination of the system geometry and the measurement of two other PK parameters at a level of precision similar
to that for $\dot\omega$, would allow us to measure the moment of inertia of a neutron star for the
first time \citep{Dam88, Wex95}.  [...] this measurement is potentially very difficult [...] The double pulsar [...] would also give insight into the nature of super-dense matter.''

In \citep{Dam07} it is written: ``It was then pointed
out \citep{Dam88} that this gives, in principle, and indirect way of measuring the moment
of inertia of neutron stars [...]. However, this can be done only if one measures,
besides\footnote{The parameter $k$ is directly related to the periastron precession $\dot\omega$.} $k$, two other PK parameters with $10^{-5}$ accuracy. A rather tall order
which will be a challenge to meet.''

Some more details are released by \citet{Kra05}: ``[...] a potential measurement
of this effect allows the moment of inertia of a neutron
star to be determined for the first time \citep{Dam88}. If two parameters, e.g. the Shapiro parameter $s$ and
the mass ratio $R$, can be measured sufficiently accurate,
an expected $\dot\omega_{\rm exp}$ can be computed from the intersection
point.''

Here we wish to examine, more precisely and with more quantitative details than done in the existing literature, the conditions which would make feasible to measure $I_{\rm A}$ at $10\%$ or better in the \psr system in view of the latest timing results \citep{Kra06}.
%Our analysis fully confirms the need of having three parameters measured with high accuracy expressed for the first time by \citep{Dam88} and %successively restated with more details by \citep{Kra05} and \citep{Dam07}.
In particular, we will show how crucial the impact of the mismodelling in the known precessional effects affecting the periastron rate of \psr is
if other effects are to be extracted from such a post-Keplerian parameter.  Such an analysis will turn out to be useful also for purposes other than measuring gravitomagnetism like, e.g., putting more severe and realistic constraints on the parameters entering various models of modified gravity. Indeed, in doing that for, e.g., a uniform cosmological constant $\Lambda$ \citet{Jet} took into account only the least-square covariance sigma of the estimated periastron rate ($6.8\times 10^{-4}$ deg yr$^{-1}$): the systematic bias due to  the first post-Newtonian (1PN) periastron precession was neglected.
\section{The systematic uncertainty in the 1PN periastron precession}
By assuming $I\approx 10^{38}$ kg m$^2$ \citep{Mor04, Bej05}, the gravitomagnetic spin-orbit periastron precession is about $\olt\approx 10^{-4}$ deg yr$^{-1}$, while the error $\delta\oms$ with which the periastron rate is phenomenologically estimated from timing data is currently $6.8\times 10^{-4}$ deg yr$^{-1}$ \citep{Kra06}. In order to measure the gravitomagnetic  effect$-$and, in principle, any other dynamical feature affecting the periastron$-$ $\delta\oms$ is certainly of primary importance, but it is not the only source of error to be carefully considered: indeed,
there are other terms contributing to the periastron precession (first and second post-Newtonian, quadrupole, spin-spin  \citep{Bar75a, Dam88, Wex95}) which must be subtracted from  $\oms$, thus  introducing a further systematic uncertainty due to the propagation of the errors in the system's parameters entering their analytical expressions. A preliminary analysis of such aspects, can be found in \citep{Lat05}. However, apart from the fact that its authors make use of the value for $i$ measured with the scintillation method \citep{Col05} which is highly uncertain for the reasons given below, in using the third Kepler law to determine the sum of the masses they also confound the relative projected semimajor axis\footnote{Our $a$ must not be confused with the one used in \citep{Lat05} which is, in fact, $a_{\rm bc}$. } $a\sin i$ (see \rfr{ohe}) with the barycentric projected semimajor axis $x$, which is the true measurable quantity from timing data, so that their analysis cannot be considered reliable.  The semimajor axis $a$ of the relative motion of A with respect to B  in a binary system is just the sum of the semimajor axes $a_{\rm bc}$ of A and B with respect to the system's barycenter, i.e.
$a = a_{\rm bc}^{\rm A} + a_{\rm bc}^{\rm B}$.

Let us, now, consider the largest contribution to the periastron rate, i.e. the 1PN precession \citep{Dam86, Dam92}
\eqi \oge = \rp{3}{(1-e^2)}\left(\rp{P_{\rm b}}{2\pi}\right)^{-5/3}(T_{\odot}M)^{2/3},\lb{oge}\eqf
where $T_{\odot}=G{\rm M}_{\odot}/c^3$ and $M=m_{\rm A}+m_{\rm B}$, in Solar masses,. It is
often referred to as gravitoelectric in the weak-field and slow-motion approximation: in the context of the Solar System it is the well known Einstein Mercury's perihelion precession of about 43 arcsec cy$^{-1}$. Thus,
\begin{equation}\left\{\begin{array}{lll}
\olt = \oms-\oge-\dot\omega_{\rm 2PN},\\\\
\delta\olt \leq \delta\oms + \delta\oge +\delta\dot\omega_{\rm 2PN}
 \lb{prece}
\end{array}\right.\end{equation}
The sum of the masses $M$  enters \rfr{oge}; as we will see, this implies that  the relative semimajor axis $a$ is required as well.
For consistency reasons,  the values of such parameters used to calculate \rfr{oge} should have been obtained independently of the periastron rate itself. We will show that, in the case of PSR J0737-3039A/B, it is possible.

Let us start from the relative semimajor axis
\eqi a=
  \left(1+R\right)\left(\rp{cx_{\rm A}}{\sin i}\right)= 8.78949386\times 10^8\ {\rm m}\lb{ohe}.\eqf
It is built in terms of $R$, the projected semimajor axis $x_{\rm A}$ and $\sin i$;  the phenomenologically estimated
post-Keplerian parameter $s$ determining the shape of the logarithmic Shapiro time delay can be identified with $\sin i$ in general relativity and  the
ratio $R=x_{\rm B}/x_{\rm A}$ has been obtained from the phenomenologically determined projected semimajor axes, being equal to the ratio of the masses in any Lorentz-invariant theory of gravity  \citep{Dam85, Dam88, Dam92} \eqi R=\rp{m_{\rm A}}{m_{\rm B}}+\mathcal{O}\left(\rp{v^4}{c^4}\right).\eqf
The uncertainty in $a$ can be conservatively evaluated as
\eqi \delta a \leq   \delta
a|_R + \delta a|_s + \delta a|_{x_{\rm A}} ,\eqf with
\begin{equation}\left\{\begin{array}{lll}
\delta a|_R\leq  \left(\rp{cx_{\rm A}}{s}\right)\delta R = 466,758\ {\rm m},\\\\
\delta a|_s\leq a\left(\rp{\delta s} s\right) = 342,879\ {\rm m},\\\\
\delta a|_{x_{\rm A}} \leq a\left(\rp{\delta x_{\rm A}}{x_{\rm A}}\right)=621 \ {\rm m}.

 \lb{erra}
\end{array}\right.\end{equation}
Thus, \eqi \delta a \leq 810,259\ {\rm m}.\lb{longo}\eqf
  \rfr{longo} yields a relative uncertainty of \eqi \rp{\delta a}{a} = 9\times 10^{-4}.\eqf It is
important to note that $x_{\rm B}$, via $R$, and $s$ have a major impact on the
overall uncertainty in $a$; our estimate has to be considered as
conservative because we adopted for $\delta s$ the largest value
quoted in \citep{Kra06}. In regard to the inclination, we did not use the more precise value for $i$ obtained from scintillation measurements in\footnote{See also \citep{Lyu05}.} \citep{Col05} because it
is inconsistent with that
derived from timing measurements \citep{Kra06}. Moreover, the scintillation method is model-dependent and it is not
only based on a number of assumptions about the interstellar medium, but it is also much
more easily affected by various other effects.   However, we will see that also $x_{\rm A}$ has a non-negligible impact.
Finally, let us note that we purposely linearly summed up the individual sources of errors in view of the  existing correlations among the various estimated parameters \citep{Kra06}.
%An expression for the relative semimajor axis  equivalent to \rfr{ohe} is
%\eqi a^{'}=(x_{\rm A}+x_{\rm B})\rp{c}{s}\lb{referee}:\eqf it is obtained from phenomenologically determined quantities %not coming from the periastron rate itself, so that it can, in principle, be used for our purposes.
%It yields a nominal value for the relative semimajor axis  which differs of only $|a-a^{'}|=1\times 10^4$ m from the %one obtained from \rfr{ohe}: since the total uncertainty $\delta a+\delta a^{'}\leq 1.63\times 10^6$ m, we can conclude %that such values are compatible each other. However, since \rfr{ohe} is slightly  more precise than \rfr{referee} by %one order of magnitude ($\delta a\leq 8.1\times 10^5$ m, $\delta a^{'}\leq 8.23\times 10^6$ m), our use of \rfr{ohe} is %fully justified.

Let us, now, determine the sum of the masses: recall that it must not come from the periastron rate itself. One possibility is to use the phenomenologically determined orbital period $P_{\rm b}$ and the third Kepler law getting\footnote{In principle, also the 1PN correction to the third Kepler law calculated by \citep{Dam86} should be included, but it does not change the error estimate presented here.}
\eqi GM= a^3\left(\rp{2\pi}{P_{\rm b}}\right)^2.\lb{Massa}\eqf
With \rfr{ohe} and \rfr{Massa} we can, now, consistently calculate \rfr{oge} getting
\eqi \oge = \rp{3}{(1-e^2)}\left(\rp{x_A+x_B}{s}\right)^2\left(\rp{2\pi}{P_{\rm b}}\right)^3=16.90410\ {\rm deg\ yr}^{-1}; \eqf
in this way the 1PN periastron precession is written in terms  of the four Keplerian parameters $P_{\rm b}, e, x_{\rm A}, x_{\rm B}$ and of the post-Keplerian parameter $s$.
The mismodeling in them yields

\begin{equation}\left\{\begin{array}{lll}
\delta \oge|_{x_{\rm B}}\leq 2\oge\left[\rp{\delta x_{\rm B}}{(x_{\rm A} + x_{\rm B} )}\right] = 0.01845\ {\rm deg\ yr}^{-1},\\\\
\delta \oge|_{s}\leq  2\oge\left(\rp{\delta s}{s}\right)= 0.01318\ {\rm deg\ yr}^{-1},\\\\
\delta \oge|_{x_{\rm A}}\leq  2\oge\left[\rp{\delta x_{\rm A}}{(x_{\rm A} + x_{\rm B} )}\right]= 1\times 10^{-5}\ {\rm deg\ yr}^{-1},\\\\
\delta \oge|_{\rm e}\leq  2e\oge\left[\rp{\delta e}{(1-e^2)}\right]= 2\times 10^{-6}\ {\rm deg\ yr}^{-1},\\\\
\delta \oge|_{P_{\rm b}}\leq  3\oge\left(\rp{\delta P_{\rm b}}{ P_{\rm b} }\right) = \mathcal{O}(10^{-8})\ {\rm deg\ yr}^{-1}.
 \lb{erra}
\end{array}\right.\end{equation}

Thus, the total uncertainty is
\eqi \delta\oge\leq 0.03165\ {\rm deg\ yr}^{-1},\lb{kazza}\eqf
which maps into a relative uncertainty of \eqi \rp{\delta\oge} \oge=1.8\times 10^{-3}.\lb{canc}\eqf
As a consequence, we have the important result
\eqi\Delta\dot\omega\equiv\oms-\oge=(-0.00463\pm 0.03233)\ {\rm deg\ yr}^{-1}.\lb{emho}\eqf
Every attempt to measure or constrain effects predicted by known Newtonian and post-Newtonian physics (like, e.g., the action of the quadrupole mass moment or the gravitomagnetic field), or by modified models of gravity, for the periastron of the \psr system must face with the bound of \rfr{emho}.

%Incidentally, we mention that we could have also used $R$ and the Shapiro time delay $r$ to separately determine the %masses of the two pulsars, but the results, in this case, would be less accurate because $M=2.6116\pm 0.1402$.
%For the periastron rate we have
%\eqi \oge = 17.13493\ {\rm deg\ yr}^{-1},\eqf
%so that
%\eqi \Delta\dot\omega\equiv\oms-\oge=(-0.23546\pm 1.41965)\ {\rm deg\ yr}^{-1},\lb{schifo}\eqf compatible with %\rfr{emho},
%but yielding a relative accuracy of only $8\times 10^{-2}$.

Should we decide to use both the post-Keplerian parameters related to the Shapiro delay  \citep{Dam86, Dam92}
\begin{equation}\left\{\begin{array}{lll}
r=T_{\odot}m_{\rm B},\\\\
s = x_{\rm A}\left(\rp{P_{\rm b}}{2\pi}\right)^{-2/3} T_{\odot}^{-1/3}M^{2/3}m_{\rm B}^{-1},
\lb{shap}
\end{array}\right.\end{equation}
for determining the sum of the masses, we would have, with \rfr{ohe},
\eqi \oge = \rp{3}{(1-e^2)}  \left( \rp{P_{\rm b}}{2\pi} \right)^{3/2} \left( \rp{r}{x_{\rm A}}  \right)^{9/4} \rp{ s^{19/4} }{(x_{\rm A}+x_{\rm B})^{5/2}},\lb{nueva}\eqf
which yields
\eqi \oge = 17.25122\pm 2.11819\ {\rm deg}\ {\rm yr}^{-1}.\eqf
The major source of uncertainty is $r$, with 2.06264 deg yr$^{-1}$; the bias due to the other parameters is about the same as in the previous case.

Let us, now, consider the second post-Newtonian contribution to the periastron precession \citep{Dam88, Wex95}
\eqi \ogge = \rp{ 3(GM)^{5/2} }{c^4 a^{7/2}(1-e^2)^2}\left\{ \left[\rp {13} 2\left(\rp{m_{\rm A}^2 + m_{\rm B}^2}{M^2}\right) +\rp {32} 3\left(\rp{m_{\rm A} m_{\rm B}}{M^2}\right)\right]\right\},\eqf up to terms of order $\mathcal{O}(e^2)$.
For our system it amounts to $4\times 10^{-4}$ deg yr$^{-1}$, so that it should be taken into account in $\Delta\dot\omega$. However, it can be shown that the bias induced by the errors in $M$ and $a$ amounts to
$4\times 10^{-6}$ deg yr$^{-1}$, thus affecting the gravitomagnetic precession at the percent level.

\section{Discussion and conclusions}
\citet{OCo04}, aware of the presence of other non-gravitomagnetic contributions to the pulsar's periastron rate,  proposed  to try to measure the  gravitomagnetic  spin-orbit precession of the orbital angular momentum \citep{Bar75a} (analogous to the Lense-Thirring node precession in the limit of a test particle orbiting a massive body) which is not affected by larger gravitoelectric contributions. However, its magnitude is $\approx (10^{-4}\ {\rm deg\ yr}^{-1})\sin\psi$, where $\psi$ is the angle between the orbital angular momentum and the pulsar's spin; thus, it would be negligible in the \psr system because of the near alignment between such vectors \citep{Sta06}, in
agreement with the observed lack of profile variations \citep{Man05,Kra06}.

In regard to the measurement of the moment of inertia of the component A via the gravitomagnetic periastron precession, our analysis has pointed out that  the bias due to the mismodelling in the 1PN gravitoelectric contribution to periastron precession--expressed in terms of the phenomenologically measured parameters $P_{\rm b}, e, x_{\rm A}, x_{\rm B}, s$--is the most important systematic error exceeding the expected gravitomagnetic rate, at present, by two orders of magnitude; the major sources of uncertainty in it are $x_{\rm B}$ and $s$, which should be measured three orders of magnitude better than now to reach the $10\%$ goal.  The projected semimajor axis $x_{\rm A}$ of A, if known one order of magnitude better than now, would induce a percent-level bias.  Instead, expressing the 1PN gravitoelectric periastron rate in terms of $P_{\rm b}, e, x_{\rm A}, x_{\rm B}, s, r$ would be definitely not competitive because the improvement required for $r$ would amount to five orders of magnitude at least. We prefer not to speculate now about the size of the improvements in timing of the \psr system which could be achieved in future.
Since the timing data of B are required as well for $x_{\rm B}$  and in view of the fact that B appears as a strong radio
source only for two intervals, each of about 10-min duration, while its pulsed emission is rather
weak or even undetectable for most of the remainder of the orbit \citep{Lyn04, Bur05}, the possibility of reaching  in a near future the required accuracy to effectively constrain $I_{\rm A}$ to $10\%$ level or better should be  considered with more skepticism than done so far. Measuring gravitomagnetism is a challenging enterprize. %
\section*{Acknowledgments}
 I gratefully thank the anonymous referees for their useful comments. I acknowledge the support from the Accademia Nazionale dei Lincei and the \"{O}sterreichische Akademie der Wissenschaften (\"{O}AW). I am grateful to H.I.M. Lichtenegger for the hospitality at the Institut f\"{u}r Weltraumforschung (\"{O}AW), Graz.
%-----------------------------------------

\newpage

\begin{table}

\caption{ Relevant Keplerian and post-Keplerian parameters of the binary system
\psr \citep{Kra06}. The orbital
period $P_{\rm b}$ is measured with a precision of $4\times 10^{-6}$ s. The projected semimajor axis is
defined as $x=(a_{\rm bc}/c)\sin i,$  where $a_{\rm bc}$ is the
barycentric semimajor axis (the relative semimajor axis $a = (x_{\rm A} + x_{\rm B})c/\sin i$), $c$ is the speed of light and $i$ is the angle
between the plane of the sky, perpendicular to the line-of-sight,
and the orbital plane. The eccentricity is $e$. The best determined post-Keplerian parameter is, to date, the periastron rate $\dot\omega$ of the component A. The phenomenologically determined post-Keplerian parameter $s$, related to the general relativistic Shapiro time delay,
is equal to $\sin i$; we  have
conservatively quoted the largest error in $s$ reported in
\citep{Kra06}.  The other post-Keplerian parameter related to the Shapiro delay, which is used in the text, is $r$.
%The
%rotational period ${\mathcal{P}_{\rm A}}=2\pi/\Omega_{\rm A}$ of
%\psr A amounts to 22 ms, while ${\mathcal{P}_{\rm
%B}}=2\pi/\Omega_{\rm B}=2.75$ s.
}
\label{tavola}

\begin{tabular}{lllllll}

\noalign{\hrule height 1.5pt}

 $P_{\rm b}$ (d)& $x_{\rm A}$ (s) & $x_{\rm B}$ (s) & $e$ & $\dot\omega$ (deg yr$^{-1}$) & $s$ & $r$ ($\mu$s)\\
\hline
$0.10225156248(5)$ & $1.415032(1)$ & $1.5161(16)$ &  $0.0877775(9)$ & $16.89947(68)$  & $0.99974(39)$  &  $6.21(33)$ \\

\hline

\noalign{\hrule height 1.5pt}

\end{tabular}

\end{table}

\end{document}